\documentstyle[prl,preprint,aps,epsf]{revtex}
\tighten
\begin{document}
\draft
\title{A small superconducting grain in the canonical ensemble}
\author{A. Mastellone$^{(1,2)}$, G. Falci$^{(1)}$ and Rosario Fazio$^{(1)}$}
\address{
 $^{(1)}$Istituto di Fisica, Universit\`a di Catania \& INFM,
         viale A. Doria 6, 95129 Catania - Italy\\
 $^{(2)}$Dipartimento di Fisica, Universit\`a "Federico II" di Napoli
        \& INFM, Mostra d'Oltremare, Pad. 19, 80125 Napoli - Italy\\
 }
\maketitle

\begin{abstract}
By means of the Lanczos method we analyze superconducting
correlations in ultrasmall grains at fixed particle number. 
We compute the ground state
properties and the excitation gap of the pairing Hamiltonian 
as a function of the level spacing $\delta$.
Both quantities turn out to be {\em parity dependent} and {\em universal}
functions of the ratio $\delta/\Delta$ ($\Delta$ is the BCS gap).
We then characterize superconductivity in the canonical
ensemble from the scaling  behavior of correlation functions in
energy space.
\end{abstract}


\narrowtext

What is the size limit for a metal particle to have superconducting
properties? Anderson~\cite{Anderson59} posed this question back in
1959 arguing that when the average level spacing $\delta $
(inversely proportional to volume of the grain) becomes of the
order of the BCS gap $\Delta$ superconductivity should disappear. 
A related question is how to characterize "superconductivity"
in small systems.
The  transition is washed out, for instance, by
thermal fluctuations of the order parameter~\cite{Muhlschlegel72}.
Moreover the hallmarks of Cooper pair condensation like the zero
resistance and the Meissner effect are absent when the grains are
of submicron size.

In a series of recent experiments  Ralph, Black and
Tinkham~\cite{Ralph95,Black96} studied the transport through
nanometer-scale Al grains.  These experiments revealed the
existence of a spectroscopic gap larger than the average level
spacing which could be driven to zero  by applying a suitable
magnetic field.  This was convincingly interpreted as the
reminiscence of superconductivity. As the grain size was further
reduced ($\,< 5\,$nm) no trace of the gap in the
spectrum was detected. The experimental results were  found to be
{\em parity dependent}, i.e. depending on the electron number in
the grain being even or odd.

In the light of these experiments von Delft et {\em al.}~\cite{vonDelft96}
reconsidered the question posed by Anderson. They included
the effect of a uniform finite level spacing  in  a parity dependent mean
field theory~\cite{Janko94}, and found that the breakdown of superconductivity
(in the BCS sense) occurs at a value of $\delta / \Delta$ which is
indeed parity dependent.
In grains with an even number of electrons superconductivity
persists down to smaller grain sizes as compared with the odd ones.
This parity effect gets enhanced when the effect of level
statistics~\cite{Smith96} is included.

Due to the spontaneous breaking of the gauge symmetry, the BCS theory is
most transparently formulated in the grancanonical
ensemble~\cite{Tinkham96}
since it is easy to define the order parameter
as the amplitude to create (or destroy) a Cooper pair in the
condensate. In the canonical ensemble this quantity vanishes 
and the characterization of coherence is more difficult.
Fortunately the use of the grancanonical ensemble is appropriate
for many systems which are large enough to give small relative
fluctuations of the electron number. This condition is not strictly
met in the experiments of Refs.~\cite{Ralph95,Black96} where
charging effects allow to fix the number of electrons in the
grain~\cite{Iachello94}. For these systems moreover quantum
fluctuations of the pairing field may become so large to invalidate
the mean field approach.
In order to characterize the ground-state pair correlations,
Matveev and Larkin~\cite{Matveev97} then proposed to use the parity
gap $\Delta_P$, an experimentally accessible quantity
related to the extra ground state energy of a system with an
unpaired electron.
Even the limit of very small grains $\Delta_P$ is very sensitive to
superconducting fluctuations (in the opposite case it reduces to
$\Delta$).

\begin{figure}
{\epsfxsize=12cm\epsfysize=10cm\epsfbox{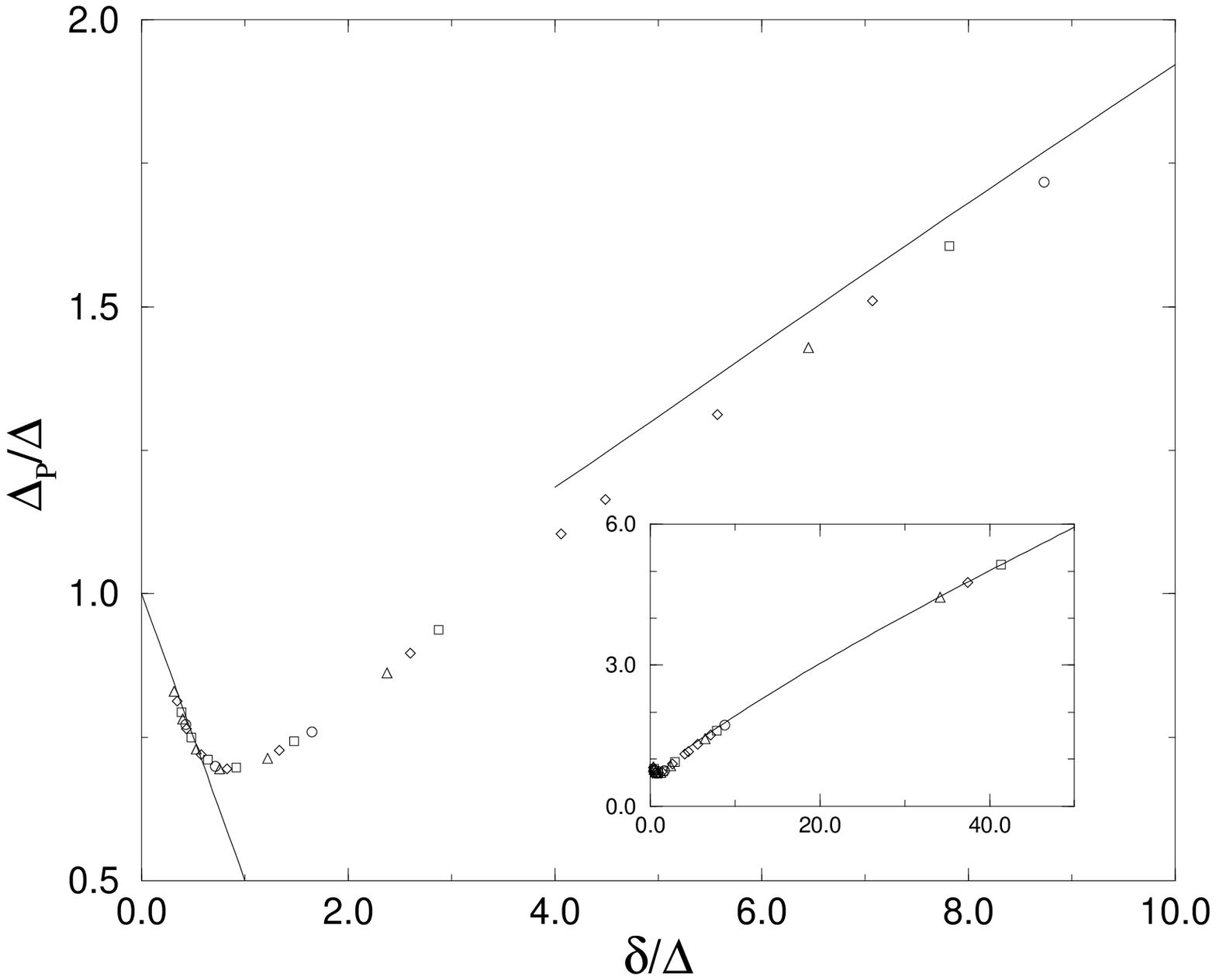}}
\caption{The parity gap is plotted as a function of $\delta/\Delta$ for
        systems with odd electron number. All the sets are evaluated at
        half filling ($\Omega=N$). In the inset the region of very large
        level spacing is plotted to compare with the asymptotic result of
        Matveev and Larkin. The parity gap is centered around $N=17$
        ($\circ$), $N=19$ ($\Diamond $),  $N=21$ ($ \Box $),
        $N=23$ ($\bigtriangleup  $).
In the limit of very small grains we got a quantitative agreement by using
$ \Delta_P \sim  \delta /2\ln (a\delta/\Delta)$ with $a\sim1.35$}
\label{fig1}
\end{figure}

\begin{figure}
{\epsfxsize=12cm\epsfysize=10cm\epsfbox{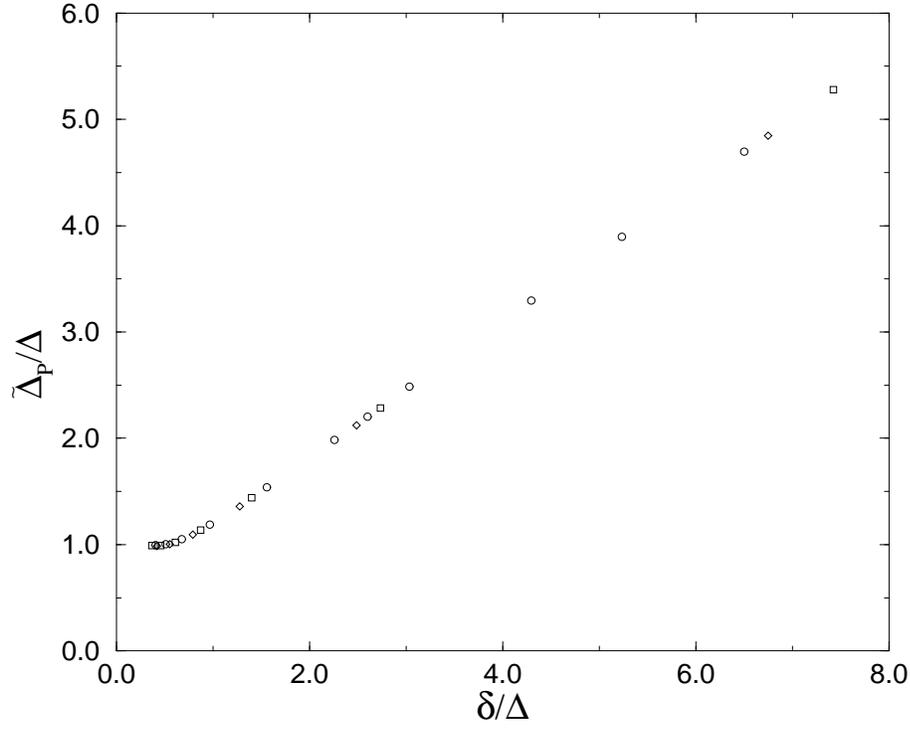}}
\caption{The same as in Fig.1 for even number of particles. The parity gap
        is centered around $N=18$ ($\circ$),  $N=20$ ($ \Box $),
        $N=22$ ($\Diamond  $).}
    \label{fig2}
\end{figure}

\begin{figure}
{\epsfxsize=12cm\epsfysize=10cm\epsfbox{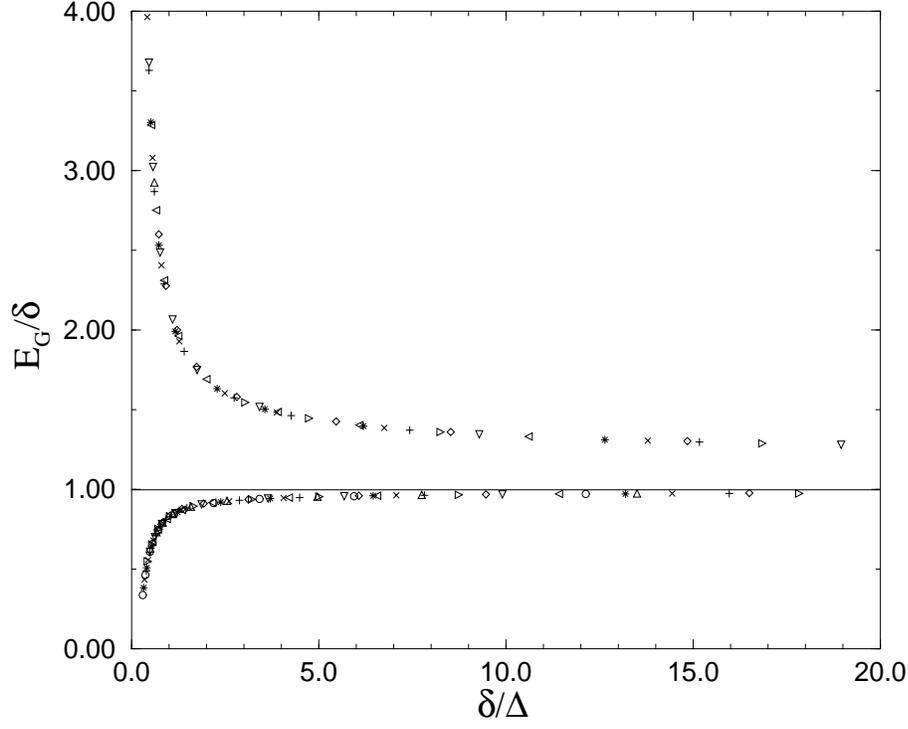}}
\caption{The spectroscopic gap $E_G$ is plotted as a function of
        $\delta/\Delta$ for even and odd particle systems at half filling.
        ($\Diamond \; \Omega =9,10$; $\bigtriangleup \; \Omega =11,12$;
        $\lhd  \; \Omega =13,14$; $\rhd  \; \Omega =15,16$;
        $\bigtriangledown  \; \Omega =17,18$; $+ \; \Omega =19,20$;
        $\times  \; \Omega =21,22$; $\star  \; \Omega =23,24$;
        $\circ  \; \Omega =25$). }
    \label{fig3}
\end{figure}

\begin{figure}
{\epsfxsize=12cm\epsfysize=10cm\epsfbox{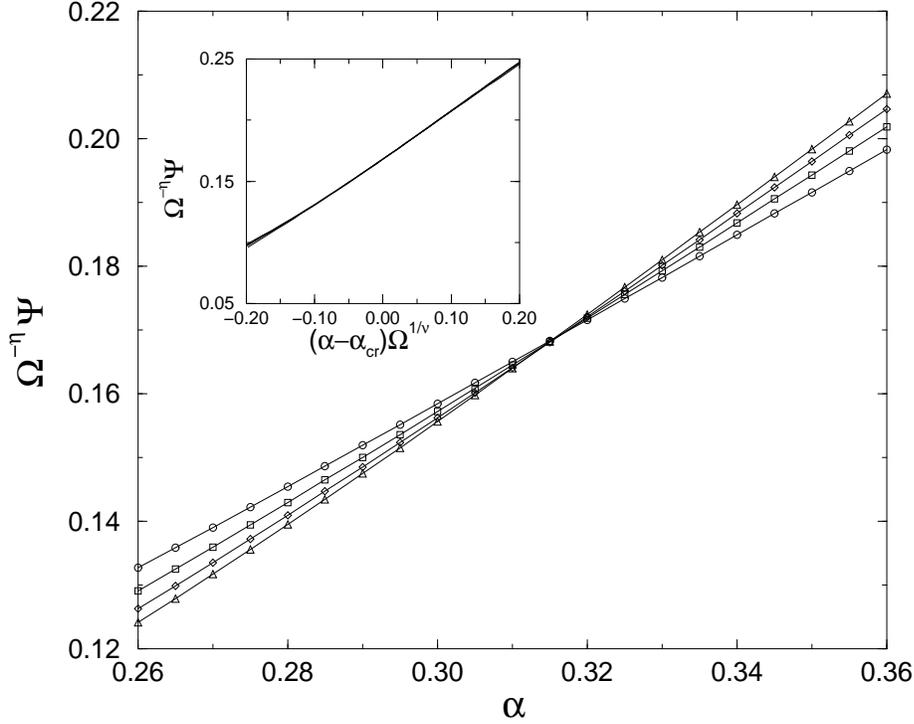}}
\caption{Data for the pair mixing parameter $\Psi$, as a function of
        $\alpha = g/\delta$ and $\Omega = N= even$, which show clearly
        a phase transition ($N=8$ ($\circ$),  $N=12$ ($\Box$),
        $N=16$ ($\Diamond$),  $N=20$ ($\bigtriangleup $)).
        In the inset it is shown the data collapse.}
    \label{fig4}
\end{figure}

\begin{figure}
{\epsfxsize=12cm\epsfysize=10cm\epsfbox{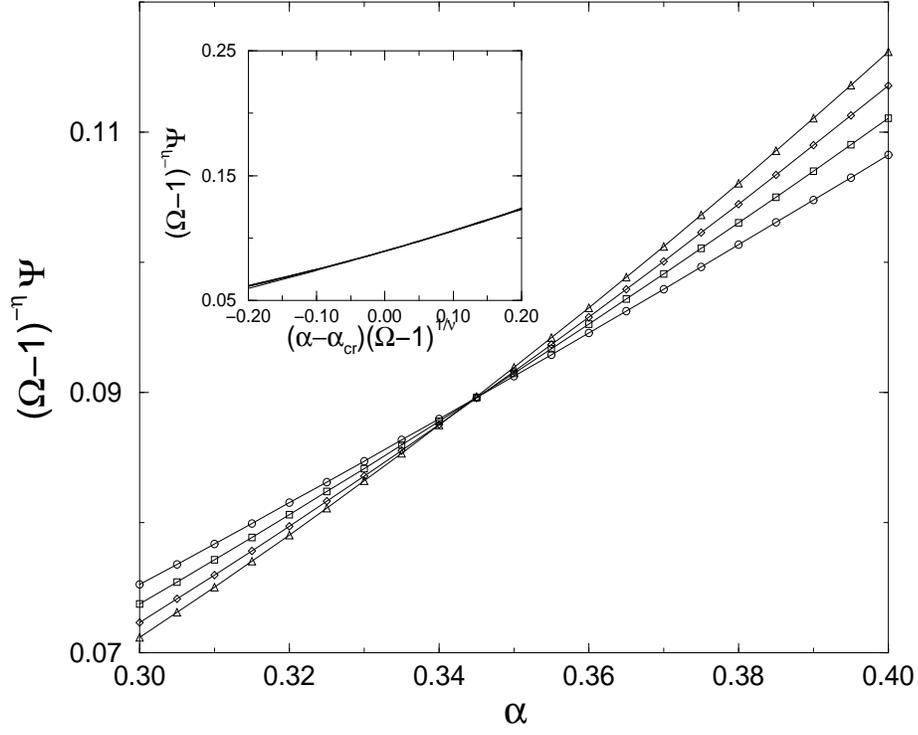}}
  \caption{The same as in Fig.4 for odd systems ($N=9$ ($\circ$),
        $N=13$ ($ \Diamond $), $N=17$ ($\Box$),
        $N=21$ ($\bigtriangleup $)). }
    \label{fig5}
\end{figure}

Superconductivity in ultrasmall grains requires to study
simultaneously the effect of finite level spacing and quantum
fluctuations at {\em fixed} particle number. In this Letter 
we tackle this problem by Lanczos exact diagonalization. We characterize the
superconducting correlations by studying the parity effect in the
ground state (Fig.\ref{fig1} and Fig.\ref{fig2}) and the
spectroscopic gap (Fig.\ref{fig3}).
Then we address the central question of defining
superconductivity at fixed particle number by
discussing the scaling properties, {\em in energy space}, of the
pairing model (see Fig.\ref{fig4} and Fig.\ref{fig5}).

The BCS pairing Hamiltonian for the small grain is
\begin{equation}
        H = \sum_{{n=1}\atop{\sigma=\pm}}^{\Omega} \epsilon_n \,
        c_{n,\sigma}^{\dagger}  c_{n,\sigma} - \alpha \, \delta \sum_{m,n=1}^{\Omega}
        c_{m,+}^{\dagger}c_{m,-}^{\dagger} c_{n,-}c_{n,+}
\label{hamiltonian}
\end{equation}
The indices $m$ and $n$ label the single particle energy levels
with energy $\epsilon_m$ and annihilation operator $c_{m,\sigma}$.
The quantum number $\sigma =\pm$  labels time reversed electron
states. The number of (doubly degenerate) levels is fixed to
$\Omega$ which is twice the Debye frequency $\omega_D$ in units of
$\delta$. Finally $\alpha$ is the dimensionless BCS coupling
constant and $\delta$ is the average level spacing ($\delta \sim
1/N(0)V$, $N(0)$ being the density of states at the Fermi energy
and $V$ the volume of the grain). Since the Hamiltonian  contains
only pairing terms, an electron in a singly occupied level cannot
interact with the other electrons (the unpaired electron is
frozen). In the following we will use the simplified model with
equally spaced single particle levels $\, \epsilon_m = \delta \, m
\,$\cite{vonDelft96,Matveev97}. We will comment later about the effect of level
statistics~\cite{Gorkov65Efetov83}. 
We study systems up to $\Omega = 25$ at
half filling ($\Omega=N$) which corresponds to the usual case of
attractive interaction in a shell $\mid \epsilon \mid < \omega_D$
centered at the Fermi energy.

We first consider the properties of the ground state by measuring
the parity gaps~\cite{Matveev97}
\begin{eqnarray}
        \Delta_{P} & = &E_{2N+1} - \frac{1}{2} \; ( E_{2N} + E_{2N+2})
\nonumber
\\
        \widetilde{ \Delta }_{P} & = &   - E_{2N} + \frac{1}{2} \;
                                ( E_{2N+1} + E_{2N-1})
\nonumber
\end{eqnarray}
Here $E_{N}$ is the ground state energy for a system with $N$ electrons.
By increasing the level spacing, $\widetilde{ \Delta }_{P}$ and $\Delta _{P}$
behave in a different way. The case $\delta \ll \Delta$ has been discussed
in~\cite{Matveev97,Golubev94,vonDelft96},
$\Delta_P/ \Delta \approx 1 - \delta/ 2 \Delta$
while $\widetilde{ \Delta }_{P}$ has only exponentially small corrections,
$
\widetilde{ \Delta }_{P}/\Delta \approx 1 - \sqrt{\delta/ \Delta} \,
\exp\{-2 \pi \delta/ \Delta \}
$. In the opposite limit, $\delta \gg \Delta$ the behavior of the
parity effect is dominated by strong superconducting
fluctuations~\cite{Matveev97} which give logarithmic corrections to
the non interacting result, i.e. $ \Delta_P = \delta
/2\ln (\delta/\Delta)$ (much larger than the BCS gap at the same
level spacing). By comparing these two limits Matveev and Larkin
concluded that a minimum should appear when the level spacing is of
the order of $\Delta$. In Fig.~\ref{fig1} the results of the
numerical diagonalization for $\Delta_P$ are presented as a
function of $\delta/\Delta$. The two asymptotic behaviors discussed
above are plotted for comparison.
In Fig.~\ref{fig2} we plot $\widetilde{\Delta}_P/\Delta$ which
monotonically increases when the grain size is reduced, as
expected~\cite{Matveev97}. Given the maximum number of levels we
can account for, in the limit $\delta \ll \Delta$ we need to
consider couplings up to $\alpha=0.5$. In this regime we use the
relation $\Delta = \omega_D /2\sinh(1/\alpha)$ to get the
comparison of our exact diagonalization results
with the asymptotics given in
Refs.~\cite{Matveev97,Golubev94,vonDelft96}. Notice that our
data, which refer to systems with different number of electrons
(from $N=10$ to $N=25$) collapse on a single curve for {\em all
values} of the ratio $\delta/\Delta$ and not only in the limiting
cases discussed in Refs.~\cite{Matveev97,Golubev94,vonDelft96}.
This suggests that $\Delta_P/\Delta$ is a
{\em universal function} of $\delta/\Delta$ (a detailed account of
this issue will be presented elsewhere~\cite{Mastellone98}). One
consequence of that is 
the systems we consider,  although
small compared with the superconducting grains used in the
experiments where $N \sim 10^{3}
- 10^{5}$, may capture all the relevant features of the
model, in particular in the interesting crossover region $\delta
\sim \Delta$.

Next we study the spectroscopic gap $E_G$ between the ground state
and the first excited many body level~\cite{Ralph95,Black96}. In
the noninteracting case $E_G=\delta$ whereas in the BCS limit
either it coincides with $2\Delta$ (even-$N$ grains) or it vanishes,
$E_G \sim \delta ^2/2\Delta$ (odd-$N$ grains).
The first excited state for even-$N$ grains belongs to the
subspace in which two unpaired electrons occupy two single-particle
levels close to the Fermi energy whereas the relevant
subspace for odd-$N$ grains is obtained by moving the unpaired
electron to the next single-particle state. In
Fig.~\ref{fig3} $E_G$ is plotted as function of the grain size. 
For small grains the effect of pairing correlations in the even case
is still observable in a rather large range of $\delta/\Delta$ for which
odd grains have already reached the asymptotic behavior $E_G=\delta$.
The crossover to the ``strong coupling'' regime
occurs at values of $\delta / \Delta$ which are different in the
odd ad in the even case and roughly agree with the mean field
critical values determined by von Delft et {\em al.} in
Ref.\cite{vonDelft96}, $\delta / \Delta  \sim 4$ (even) and $\delta
/ \Delta \sim 1$ (odd). 
It is interesting to notice that the BCS regime is reached at  
(parity dependent) values of $\delta / \Delta \ll 1$ so there is an
intermediate region of values 
$\delta / \Delta \leq 1$ where
BCS theory describes well enough only the ground state 
properties~\cite{Mastellone98}.
Finally we stress that also for $E_G$ an universal behavior is found.

To summarize this first part we have shown 
the full crossover between a "weak coupling" regime (very small grains,
where fluctuational superconductivity manifests itself via logarithmic
renormalizations) and a "strong coupling" regime (very large grains).
This situation is reminiscent of the antiferromagnetic Kondo problem. 
The level spacing $\delta$ is the low energy cutoff which tunes the
system through the two regimes.  The breakdown of the logarithmic 
renormalization marks the crossover to the superconducting phase.
This provides a {\em quantitative} answer to Anderson's 
question~\cite{Anderson59} 
despite of the fact that we have bypassed the very problem of defining 
superconductivity.
Following the conventional wisdom up to now we meant by 
superconductivity a regime in which BCS results are qualitatively 
valid. This is not satisfactory in the canonical ensemble since the 
central quantity, the BCS order parameter, is always zero. 
In order to characterize superconductivity one has to consider
higher order correlators. However they will be non zero for 
generic interaction, even for repulsive ones, so  
it is not straightforward to extract from them a quantity which plays 
the role of the "order parameter".

Nevertheless a characterization of superconductivity in the canonical
ensemble can be achieved by studying the scaling of correlations in the
energy space: {\em a superconducting system displays long range energy 
correlations}. To this end we consider the pseudospin
representation~\cite{Anderson58} of  the Hamiltonian Eq.(\ref{hamiltonian})
\begin{equation}
        H = \sum_{n=1}^{\Omega} \epsilon_n (1 - 2S^z_n)
         - \alpha \, \delta \, \sum_{m,n=1}^{\Omega}S^+_mS^-_n
\label{pseudohamiltonian}
\end{equation}
where $S^+_n = c_{n,+}^{\dagger}c_{n,-}^{\dagger}$ and
$S^z_n = (1/2)(1-c_{n,+}^{\dagger}  c_{n,+} - c_{n,-}^{\dagger} c_{n,-})$.
Each energy level is represented by a site of a
fictitious lattice and Eq.(\ref{pseudohamiltonian}) is
the Hamiltonian of a one dimensional spin-1/2 XY model with long range
interaction in a nonuniform transverse field~\cite{Altshuler97}.
The number of pairs fixes the total $S^z$ and
for odd electron numbers one should simply remove the ``site'' occupied
by the unpaired electron.
In the absence of XY interaction the spins
point in the z-direction with a domain wall at the Fermi energy separating
up and down spin regions. In the opposite limit (vanishing transverse field)
the system possesses long range order in the XY plane~\cite{Kleinert78}.
The superconducting properties can be studied
by defining the appropriate correlation functions
in this fictitious lattice (see also Ref.~\cite{Altshuler96}).
Following Ref.\cite{vonDelft96,Braun98} we consider the quantity
\begin{equation}
\Psi =  \sum_{n=1}^{\Omega} u_n v_n
\label{Psi}
\end{equation}
where $v_n = \langle S^+_nS^-_n\rangle^{1/2} $ and $u_n =\langle
S^-_nS^+_n\rangle^{1/2}$. In the noninteracting case $\Psi =0$
since both $u_n$ and $v_n$ are step function symmetric around the
Fermi energy and hence their product is zero. In the limit of very
strong interaction the occupation probability for the pairs, as a
function of the level position, is roughly uniform. In this case an
estimate of the energy of this configuration, at half filling, is
$
         \sim (1/2 -\alpha) \, \delta \, \Omega^2
$.
Note that for  $\alpha > 1/2$ the system gains energy, due to pair
mixing, from arbitrary high energy levels. 
So if we enlarge the phase space available for coherence (for instance
by progressively increasing $\Omega$ at fixed $N/\Omega$) 
a ``normal'' system will not take advantage from the presence of extra levels 
whereas a ``superconducting'' does. In other words 
correlations are short ranged in energy in ``normal'' systems 
whereas a ``superconducting'' system displays long range energy correlations.

The finite size scaling Ansatz for $\Psi$ is
\begin{equation}
\Psi = \Omega^{\eta} F((\alpha - \alpha_{cr})\Omega^{1/\nu})
\label{scalingPsi}
\end{equation}
where $\alpha_{cr}$ is the critical point.
In Fig~\ref{fig4} and Fig~\ref{fig5} the behavior of $\Psi$
is shown for the even and the odd case respectively.
A scale invariant point is found whose value is
parity dependent.
By collapsing all the data on a single curve, shown for the even and the odd
case  in the inset of
Fig~\ref{fig4} and Fig~\ref{fig5} respectively, it is possible to determine the
exponent $\nu$.
We obtain $\alpha_{cr}= 0.315 \pm 0.002$, $\eta = 0.94$, $1/\nu =0.26$ for the
even case and $\alpha_{cr}= 0.345 \pm 0.002$, $\eta = 1.08$, $1/\nu =0.35$ for
the odd case.
Both the magnitude of $\Psi$ and the location of the critical point
(see Fig~\ref{fig4} and Fig~\ref{fig5}) confirm the natural
conjecture that superconducting correlations
is destroyed easier in the odd rather than in the even case.

We propose that the scaling in energy space of properly defined
correlation functions (like that defined in  Eq.(\ref{Psi})) can
characterize quantitatively superconductivity in the canonical
ensemble.

The results presented here for the parity gap and  the excitation gap
are in a good agreement with the analytical expressions of
Refs.~\cite{Matveev97,Golubev94}. If the approximation of equally
spaced levels is relaxed, mesoscopic fluctuations are expected to
be important in the intermediate region $\delta \sim
\Delta$~\cite{Matveev97}. Nevertheless the very existence of the quantum
phase transition and the scaling in energy space is not questioned
since it does depend only on the interplay between kinetic
energy and pairing interaction. The location of the critical point and the
values of the exponents will be different.

The inclusion of the level statistics as well as the role of an applied
magnetic field~\cite{Aleiner97,Braun97,Braun98} will be the subject of a
forthcoming publication~\cite{Mastellone98}.

{\bf Acknowledgments} We thank L. Amico, F. Braun, J. von Delft, K.A. Matveev,
A. Muramatsu, G. Santoro, G. Sch\"on, A. Tagliacozzo and U. Weiss
for fruitful discussions and/or comments. We thank G. Giaquinta for useful
conversations and constant support. 
We acknowledge the financial support of INFM under the PRA-QTMD,
EU TMR programme (Contract no. FMRX-CT 960042).
GF acknowledges the Institut f\"ur
Theoretische Physik-II, Universit\"at Stuttgart (D) for
hospitality.

\newpage
\newpage
\newpage
\newpage

\end{document}